\documentclass[12pt]{article}
\usepackage{a4wide}
\usepackage{epsfig}

\begin{document}
\begin{flushright}
MZ-TH/98-12\\
hep-ph/9805224\\
March 1998\\
\end{flushright}
\vspace{0.5cm}
\begin{center}
{\Large\bf A new technique for computing the}\\[7pt]
{\Large\bf spectral density of sunset-type diagrams:}\\[7pt]
{\Large\bf integral transformation in configuration space}\\[1truecm]
{\large S.~Groote,$^1$ J.G.~K\"orner$^1$ and A.A.~Pivovarov$^{1,2}$}\\[.7cm]
$^1$ Institut f\"ur Physik, Johannes-Gutenberg-Universit\"at,\\[.3truecm]
  Staudinger Weg 7, D-55099 Mainz, Germany\\[.7truecm]
$^2$ Institute for Nuclear Research of the\\[.3truecm]
  Russian Academy of Sciences, Moscow 117312
\vspace{1truecm}
\end{center}

\begin{abstract}
We present a new method to investigate a class of diagrams which generalizes 
the sunset topology to any number of massive internal lines. Our attention 
is focused on the computation of the spectral density of these diagrams 
which is related to many-body phase space in $D$ dimensional space-time.
The spectral density is determined by the inverse $K$-transform of the 
product of propagators in configuration space. The inverse $K$-transform 
reduces to the inverse Laplace transform in any odd number of space-time 
dimensions for which we present an explicit analytical result.
\end{abstract}

\newpage\noindent
Recently there has been renewed interest in the calculation of 
sunset-type diagrams with massive internal lines (as some of the latest 
refs.\ see for example~\cite{Gasser,PS,PT,davyd}). The simplicity of the 
sunset-type topology has a special appeal since sunset-type diagrams can 
be regarded as a laboratory for testing new methods in the computation of
diagrams with massive particles (see e.g.~\cite{ChKuhnKw}). For 
phenomenological applications one needs to avail of efficient numerical 
algorithms to compute the sunset-type diagrams (see e.g.~\cite{Xloops}).
In~\cite{GKP} we have derived a one-dimensional integral representation 
for a class of diagrams that generalizes the sunset topology to any number 
of internal lines (we call them water melon diagrams). We believe that the 
method described in~\cite{GKP} and in the present paper completely solves 
the problem of calculating the class of sunset-type diagrams. 

In this note we present a new representation for the spectral density of 
water melon diagrams through an integral transformation in configuration 
space. It is given by an inverse $K$-transform of the product of
propagators of internal lines that constitutes the water melon diagram.
The representation simplifies to the inverse of the standard Laplace 
transform in any odd number of space-time dimensions for which we present 
a concise analytical result. The spectral density of water melon diagrams 
is related to the many-body phase space integral in $D$ dimensional 
space-time for which we find novel results.

\begin{figure}\begin{center}
\epsfig{figure=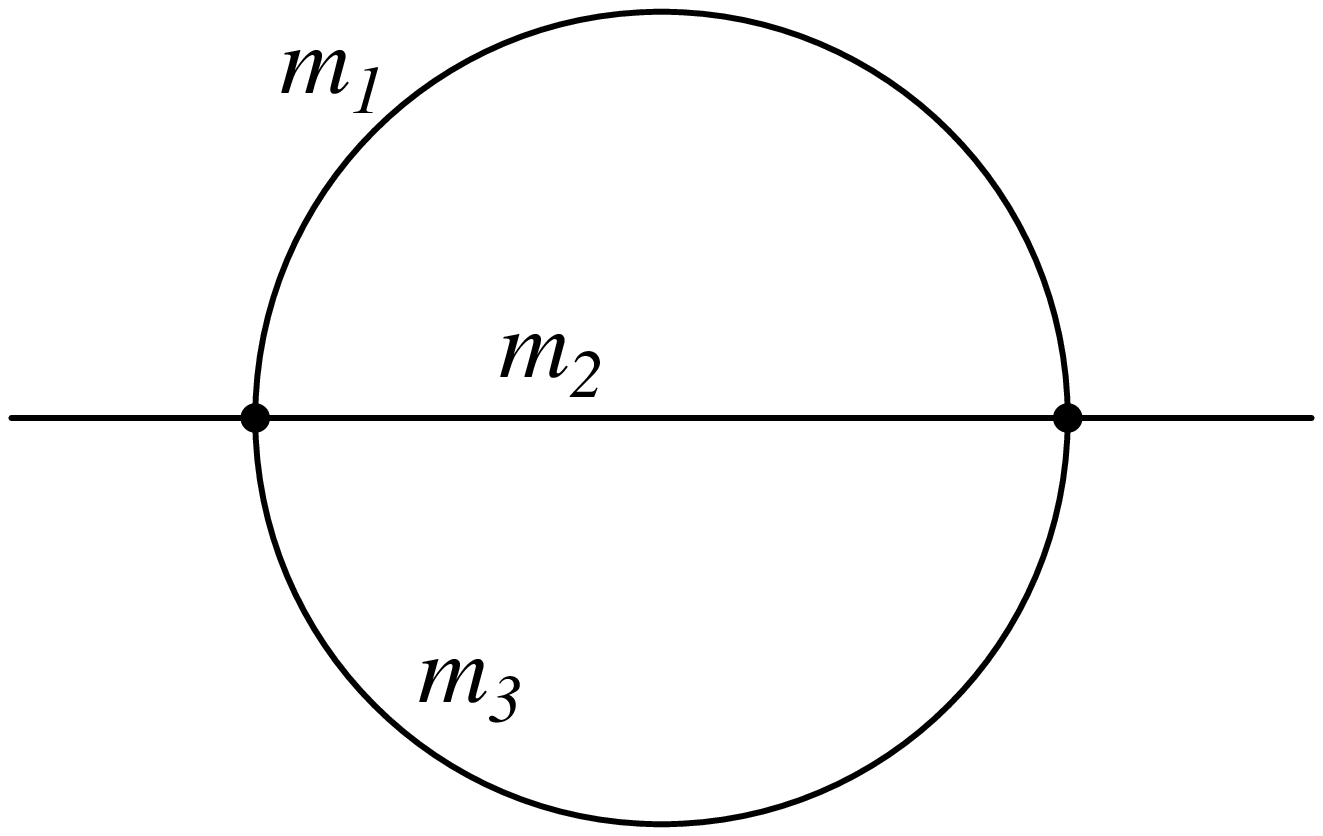, height=3truecm, width=5truecm}\kern1truecm
\epsfig{figure=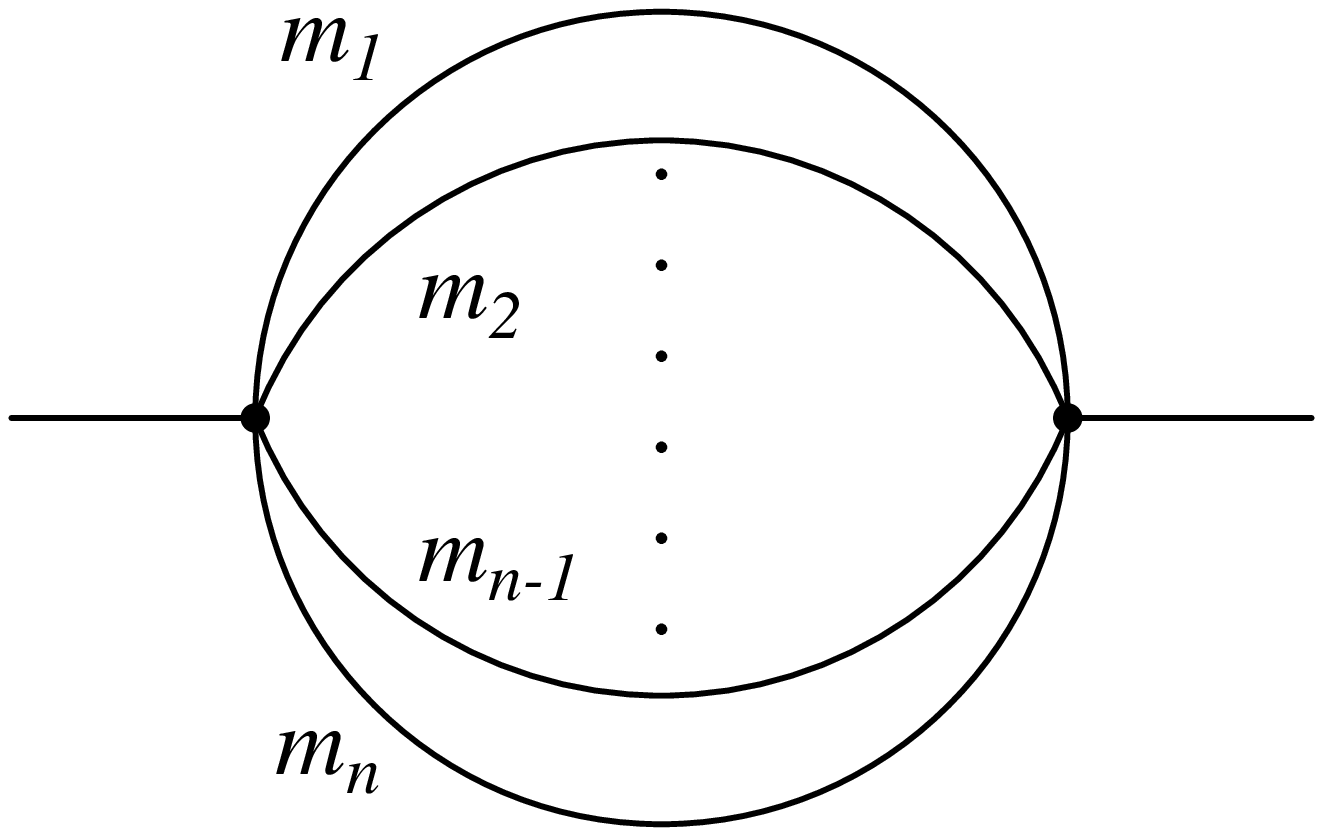, height=3truecm, width=5truecm}\vspace{12pt}
\hbox{\bf\kern4truecm(a)\kern5truecm(b)\kern4truecm}
\caption{\label{fig1}(a) a sunset diagram with three different masses $m_1$,
$m_2$ and $m_3$\hfil\break
(b) general topology of the class of water melon diagrams}
\end{center}\end{figure}

The sunset diagram proper represents the leading order perturbation theory 
correction to the propagator in $\phi^4$-theory, i.e.\ it is a two-loop 
insertion with three internal lines (see Fig.~1(a)). A straightforward 
generalization of this topology (water melon diagram) is a correction to 
the propagator in the $\phi^{n+2}$-theory that contains $n$ loops and $n+1$ 
internal lines (see Fig.~1(b)). As an example take the leading quantum 
corrections of higher order chiral perturbation theory for pseudoscalar 
mesons where such diagrams occur. Thus we describe as a generic case of 
the water melon diagram the correlator of two field monomials 
$j_n(x)=:\phi_{1,\mu_1}\ldots\phi_{n,\mu_n}\!\!:$ taken in the normal 
ordering form where $\phi_{1,\mu}$ is a derivative of the field $\phi_1$
with a multi index $\mu=\{\mu_1,\ldots,\mu_p\}$. We leave aside trivial 
generalizations that would result from the inclusion of derivatives and 
the emission/absorption of additional external particles with vanishing 
momentum. These would merely result in the doubling of the propagators to
which they are attached and thus are easily included in our general 
approach. Returning to the scalar case the general representation of the
scalar polarization function corresponding to water melon diagrams is
given by
\begin{equation}\label{deriv}
\Pi(x)=\prod_{i=1}^nD(x,m_i)
\end{equation}
where $D(x,m)$ is the configuration space propagator of a particle with 
mass $m$ in $D$-dimensional (Euclidean) space-time~\cite{GR}
\begin{equation}\label{prop}
D(x,m)=\frac1{(2\pi)^D}\int{e^{ip_\mu x^\mu}d^Dp\over p_\mu p^\mu+m^2}
  ={(mx)^\lambda K_\lambda(mx)\over(2\pi)^{\lambda+1}x^{2\lambda}}
\end{equation}
where $D=2\lambda+2$, and $K_\lambda(z)$ is a McDonald function (a 
modified Bessel function of the third kind, see e.g.~\cite{Watson}). When 
we refer to a specific space-time dimension $D$ later on we add the 
space-time index to the propagator function, i.e.\ we write $D_D(x,m)$ for 
the propagator function. The propagator depends only on the length $x$ of 
the $D$-dimensional vector $(x^\mu)$, $x=\sqrt{x_\mu x^\mu}$ (a times we 
shall also use the notation $x$ for the vector itself).

The representation Eq.~(\ref{deriv}) contains the complete information about
the class of water melon diagrams and is the starting point for all further 
investigations.  

For some applications one needs the Fourier transform of the polarization 
function which is defined by
\begin{equation}\label{cor}
\tilde\Pi(p)=\int\Pi(x)e^{ip_\mu x^\mu}d^Dx
  =\int\langle Tj_n(x)j_m(0)\rangle e^{ip_\mu x^\mu}d^Dx.
\end{equation}
The analytic structure of the Fourier transform $\tilde\Pi(p)$ is 
determined by the K\"allen-Lehmann representation
\begin{equation}
  \label{disp}
\tilde \Pi(p)=\int_0^\infty {\rho(s)ds\over s+p^2}-{\rm(subtractions)}
\end{equation}
with the appropriate number of subtractions for the regularization of 
ultraviolet divergences, if necessary. The spectral density $\rho(s)$ or 
the discontinuity across the cut in the complex plane of the analytic 
function $\tilde\Pi(p)$ is of interest for various physics applications.
We mention, though, that in the Euclidean region the function
$\tilde\Pi(p)$ itself also has some applications. For these applications
the values of $\tilde\Pi(p)$ can easily be found from the representation 
Eq.~(\ref{disp}) or directly from Eq.~(\ref{cor})~\cite{GKP}. Here we 
shall focus our attention on the spectral decomposition of the water melon 
diagrams which in turn is connected with the particle content of the 
underlying theory.

The analytic structure of the correlator $\Pi(x)$ (or the spectral density 
of the corresponding polarization operator) can be determined directly in 
configuration space without having to compute its Fourier transform first.
The dispersion representation (or the spectral decomposition) of the 
polarization function in configuration space has the form 
\begin{eqnarray}\label{xone}
\Pi(x)=\int_0^\infty\rho(s)D(x,s)ds
\end{eqnarray}
where $D(x,s)$ is the scalar propagator with the mass $m=\sqrt{s}$. This 
representation was used for sum rules applications in~\cite{Sch,fact} 
where the spectral density for the two-loop sunset diagram was found in
two-dimensional space-time~\cite{Sch1}. With the explicit form of the 
propagator in configuration space Eq.~(\ref{prop}) the representation 
in Eq.~(\ref{xone}) turns into a particular example of the Hankel 
transform, namely the $K$-transform~\cite{Ktr0,Ktr}. Up to inessential 
factors of $x$ and $\sqrt{s}$, Eq.~(\ref{xone}) reduces to the generic form 
of the $K$-transform for a conjugate pair of functions $f$ and $g$,
\begin{eqnarray}\label{ktrdef}
g(y)=\int_0^\infty f(x)K_\nu(xy)\sqrt{xy}\,dx.
\end{eqnarray}
The inverse of this transform is known to be given by 
\begin{eqnarray}\label{inv}
f(x)=\frac1{\pi i}\int_{c-i\infty}^{c+i\infty}g(y)I_\nu(xy)\sqrt{xy}\,dy
\end{eqnarray}
where $I_\nu(x)$ is a modified Bessel function of the first kind and the 
integration runs along a vertical contour in the complex plane to the right 
of the right-most singularity 
of the function $g(y)$. In order to obtain a representation for the spectral
density $\rho(s)$ of water melon diagrams in general $D$-dimensional
space-time one needs to apply the inverse $K$-transform to the particular
case Eq.~(\ref{xone}). One has 
\begin{eqnarray}\label{expDKinv}
m^\lambda\rho(m^2)={(2\pi)^\lambda\over i}
  \int_{c-i\infty}^{c+i\infty}\Pi(x)x^{\lambda+1}I_\lambda(mx)dx.
\end{eqnarray}
The transform Eq.~(\ref{expDKinv}) solves the problem of determinating 
the spectral density of water melon diagrams by reducing it to the 
computation of a one-dimensional integral for a general water melon diagram 
with any number of internal lines and different masses. Below we discuss 
some technicalities concerning the efficient evaluation of the contour
integral.

The analytic structure of the correlator in Eq.~(\ref{deriv}) is now 
explicit using the representation given by Eq.~(\ref{expDKinv}) and 
exhibits the distribution nature of the spectral density $\rho(s)$. For 
odd-dimensional space-time the representation in Eq.~(\ref{xone}) reduces
to the Laplace transformation. To obtain the spectral density (the function 
$f(x)$ in this particular example) one can use Eq.~(\ref{inv}).
For energies below threshold it is possible to close the contour of 
integration to the right. With the appropiate choice of the constant $c$ 
as specified above the closed contour integration gives zero due to the 
absence of singularities in the relevant domain of the right semi-plane. 
By closing the contour of integration to the left and keeping only that 
part of the function $I_\nu(z)$ which is exponentially falling for 
${\rm Re}(z)<0$ one can obtain another convenient integral representation 
for the spectral density when the energy is above threshold. The only 
singularities within the closed contour are then poles at the origin (in 
odd-dimensional space-time) and the evaluation of the integral can be done 
by determining the corresponding residues. These are purely algebraic 
manipulations, the simplicity of which also explain the simplicity of the 
computations in odd-dimensional space time. For a small number of internal 
lines $n$ the spectral density can also be found by using the convolution 
formulae for the spectral densities of a smaller number of particles (see 
e.g.~\cite{NarPiv}). For large $n$ the computations described 
in~\cite{NarPiv} become quite cumbersome and the technique suggested in 
the present paper is much more convenient. For example, in odd-dimensional 
space-time our techniques give the result for water melon diagrams in terms 
of elementary functions. In three space-time dimensions our results can be 
used to compute
the phase-space integrals for particles in jets where the 
momentum along the direction of the jet is fixed~\cite{Mirkes}. Another 
application can be found in three dimensional QCD which emerges as the high 
temperature limit of the ordinary theory of strong interactions for the 
quark-gluon plasma (see e.g.~\cite{qcd30,qcd3,Hatsuda,Raj}).

For even-dimensional space-time the analytic structure of $\Pi(x)$ in 
Eq.~(\ref{deriv}) is more complicated: there is a cut along the negative 
semi-axis which prevents a straightforward evaluation by simply closing 
the contour of integration to the left. The discontinuity along the cut is, 
however, well known and includes only Bessel functions that appear in the 
product of propagators for the polarization function. Therefore the 
representation~(\ref{expDKinv}) is essentially equivalent to the direct
analytic continuation of the Fourier transform~\cite{GKP} but may be more 
convenient for numerical treatment because there is no oscillating 
integrand in~(\ref{expDKinv}).
 
Note also that the expression given by Eq.~(\ref{deriv}) 
can have non-integrable
singularities at small $x$ for a sufficiently large number of 
propagators when $D>2$~\cite{BS}. Therefore the computation of its Fourier 
transform requires regularization or subtractions~\cite{GKP}. The spectral 
density itself is finite (the structure of the water melon diagrams is very 
simple and there are no subdivergences when one employs a properly defined
$R$-operation~\cite{BS}) and thus requires no regularization. The explicit 
representation for the spectral density within the more traditional 
approach reads~\cite{GKP}
\begin{eqnarray}\label{disc}
\lefteqn{2\pi i{\rm\,Disc\,}\tilde\Pi(p)|_{p^2=-s}}\nonumber\\
&=&\int_0^\infty\left(\frac{2\pi\xi}s\right)^{\lambda+1}J_\lambda(\xi)\Bigg[
  e^{-i\pi(\lambda+1)}\prod_{i=1}^n\frac14\left(\frac{m_i\sqrt s}{2\pi\xi}
  \right)^\lambda e^{i\pi(\lambda+1/2)}H^{(1)}_\lambda
  \left(\frac{m_i\xi}{\sqrt s}\right)\nonumber\\&&\qquad\qquad
  -e^{i\pi(\lambda+1)}\prod_{i=1}^n\frac14\left(\frac{m_i\sqrt s}{2\pi\xi}
  \right)^\lambda e^{-i\pi(\lambda+1/2)}H^{(2)}_\lambda
  \left(\frac{m_i\xi}{\sqrt s}\right)\Bigg]d\xi,
\end{eqnarray}
where one has taken the discontinuity of the Fourier transform across the
physical cut for $p^2=-s\pm i0$, $s>0$. The analytic continuation is done 
according to~\cite{GR}
\begin{equation}\label{hank}
K_\lambda(z)={\pi i\over 2}e^{\frac\pi2\lambda i}H^{(1)}_\lambda(iz)   
\end{equation}
with $H^{(1,2)}_\lambda(z)$ being the Hankel functions,
$H^{(1)}_\lambda(z)=(H^{(2)}_\lambda(z))^*$ for real $z$ and $\lambda$. 
This is an alternative representation of the spectral density, and in some 
instances this representation can be more convenient for numerical treatment.

In the following we present some explicit examples of applying the 
technique of computing the spectral density of water melon diagrams on the
basis of integral transformations in configuration space.

First a remark about the mass degenerate one-loop case is in order. All 
necessary integrals (both for the direct and the inverse $K$-transform) 
involve no more than the product of three Bessel functions which can be 
found in a standard collection of formulas for special functions (see 
e.g.~\cite{GR}). The spectral density in $D$-dimensional space-time (for 
two internal lines with equal masses) can be computed to be
\begin{equation}\label{specdeg}
\rho(s)=\frac{(s-4m^2)^{\lambda-1/2}}{2^{4\lambda+1}\pi^{\lambda+\frac12}
  \Gamma(\lambda+\frac12)\sqrt s},\qquad s>4m^2.
\end{equation}
This formula is useful since it can be used to test the limiting cases of 
more general results. Similarly one can derive a general $D$-dimensional
analytical result for the one-loop nondegenerate mass case.

For three-dimensional space-time we have the dispersion representation
\begin{eqnarray}\label{lap3}
\Pi(x)=\int_0^\infty\rho(m^2)D_3(x,m)dm^2
  =\int_0^\infty\rho(m^2){e^{-mx}\over 4\pi x}dm^2
\end{eqnarray}
with the three-dimensional scalar propagator
\begin{eqnarray}\label{3prop}
D(x,m)={\sqrt{mx}K_{\frac12}(mx)\over(2\pi)^{3/2}x}={e^{-mx}\over 4\pi x}.
\end{eqnarray}
One can invert Eq.~(\ref{lap3}) and obtains
\begin{eqnarray}\label{inv3}
2m\rho(m^2)=\frac1{2\pi i}\int_{c-i\infty}^{c+i\infty}4\pi x\Pi(x)e^{mx}dx
\end{eqnarray}
which is a special case of Eqs.~(\ref{inv}) and~(\ref{expDKinv}) with 
\begin{equation}
I_{\frac12}(z)=\sqrt{\frac2{\pi z}}\sinh(z)
\end{equation}
where one only needs to retain the $e^z$ piece in the hyperbolic sine 
function.
The solution given by 
Eq.~(\ref{inv3}) has the appropriate support as a distribution or, 
equivalently, as an inverse Laplace transform. It vanishes for 
$m<M=\sum_{i=1}^nm_i$ since the contour of integration can be closed to 
the right where there are no singularities of the integrand. Recall that 
for large $x$ with ${\rm Re}(x)>0$ the asymptotic behaviour of the 
polarization function $\Pi(x)$ is governed by the sum of the masses of the 
propagators and reads
\begin{eqnarray}\label{asym3}
\Pi(x)\sim\exp{(-xM)}.
\end{eqnarray}
For $m>M$ one can close the contour to the left and then the only 
singularities of $\Pi(x)$ are the poles at the origin of $\Pi(x)$ since it 
is a product of the propagators of the form of Eqs.~(\ref{deriv}) 
and~(\ref{3prop}). The integration in Eq.~(\ref{inv3}) then reduces to 
finding the residues. Indeed,
\begin{eqnarray}\label{expex3}
\Pi(x)=\prod_{i=1}^nD_3(x,m_i)={e^{-Mx}\over(4\pi x)^n},
\end{eqnarray}
and Eq.~(\ref{inv3}) gives
\begin{eqnarray}\label{3final}
2m\rho(m^2)=\frac1{2\pi i}\int_{c-i\infty}^{c+i\infty}4\pi x\Pi(x)e^{mx}dx
  =\frac1{2\pi i(4\pi)^{n-1}}\int_{c-i\infty}^{c+i\infty}
  {e^{(m-M)x}\over x^{n-1}}dx.
\end{eqnarray}
At $m>M$ one closes the contour of integration to the left and computes 
the residue at the origin to obtain ($n>1$)
\begin{equation}\label{rho3fin}
2m\rho(m^2)={(m-M)^{n-2}\over(4\pi)^{n-1}(n-2)!}\theta(m-M). 
\end{equation}
This also explains the simplicity of the structure of the spectral density 
in odd numbers of dimensions of space-time when traditional means are 
used~\cite{GKP}. In five-dimensional space-time Eq.~(\ref{inv3})
is applicable almost without any change because the propagator now reads
\begin{equation}\label{5prop}
D_5(x,m)={{(mx)}^\frac32K_{\frac32}(mx)\over(2\pi)^{5/2}x^3}
  ={e^{-mx}\over8\pi^2x^3}(1+mx).
\end{equation}
Compared to the three-dimensional case the only additional complication 
is that the degree of the order of the pole at the origin is changed and
that one now has a linear combination of terms instead of the simple monomial
in three dimensions.

In even number of dimensions one is dealing with a genuine $K$-transform.
We discuss in some detail the important case of four-dimensional space-time. 
For $D=4$ ($\lambda=1$), Eqs.~(\ref{prop}) and~(\ref{xone}) give
\begin{eqnarray}\label{K4}
\Pi(x)=\int\rho(m^2)D_4(x,m)dm^2=\int\rho(m^2)
  {mx K_1(mx)\over 4\pi^2x^2}dm^2,
\end{eqnarray}
and Eq.~(\ref{expDKinv}) is written as
\begin{eqnarray}\label{invK4}
2m\rho(m^2)=\frac1{\pi i}\int_{c-i\infty}^{c+i\infty}
  4\pi^2x^2\Pi(x)I_1(mx)dx.
\end{eqnarray}
All remarks about the behaviour at large $x$ apply here as well. However, 
the structure of singularities is more complicated than in the 
odd-dimensional case. In addition to the poles at the origin there is a cut 
along the negative semi-axis that renders the computation of the spectral 
density more involved. The cut arises from the presence of the 
functions $K_1(m_ix)$ in the polarization function $\Pi(x)$. Also the 
asymptotic behaviour of the function $I_1(z)$ is more complicated than that 
of $I_{1/2}(z)$. In particular the extraction of the exponentially falling 
component on the negative real axis is rather tricky. Incidentally, the
fall-off behaviour of the function $I_1(z)$ on the negative real axis can
be taken as an illustration of Stokes' phenomenon of asymptotic expansions 
(see e.g.~\cite{Watson}). While the analytic structure of the 
representation is quite transparent and the integration can be performed 
along a contour in the complex plane, there are some subtleties when one 
wants to obtain a convenient form for numerical treatment~\cite{GKP}.

After closing the contour to the left (for $m>M$) using the appropriate 
part of the function $I_1(z)$ we obtain
\begin{eqnarray}\label{left4}
\lefteqn{i\pi\int_{c-i\infty}^{c+i\infty}x^2\Pi(x)I_1(mx)dx}\nonumber\\ 
  &=&-\int_\epsilon^\infty K_1(mr)r^2\left(\Pi(e^{i\pi}r)+\Pi(e^{-i\pi}r)
  \right)dr+2\int_\epsilon^\infty K_1(mr)r^2\Pi(r)dr\\&&
  +\int_{C_-}(i\pi I_1(mz)+K_1(mz))\Pi(z)z^2dz
  +\int_{C_+}(i\pi I_1(mz)-K_1(mz))\Pi(z)z^2dz\nonumber
\end{eqnarray}
where the contours $C_+$ and $C_-$ are semi-circles of radius $\epsilon$
around the origin in the upper and lower complex semi-plane, respectively.
For practical evaluations of $\Pi(e^{\pm i\pi}r)$ the following rule for 
the analytic continuation of McDonald functions is used:
\begin{eqnarray}\label{kanal}
K_1(e^{\pm i\pi}mr)=-K_1(mr)\mp i\pi I_1(mr),\qquad mr>0.   
\end{eqnarray}

Let us add a few remarks on the final representation Eq.~(\ref{left4}) 
which is in a suitable form for numerical integration. We have 
introduced an auxiliary regularization in terms of a circle of finite 
radius $\epsilon$ which runs around the origin with its pole-type 
singularities. The spectral density is independent of $\epsilon$, and the 
parameter $\epsilon$ completely cancels in the full expression for the 
spectral density as given by Eq.~(\ref{left4}). This must be so since the 
spectral density is finite for the water melon class of diagrams. 
Eq.~(\ref{left4}) contains no oscillatory integrands (cf.\ 
Eq.~(\ref{disc})), and the integration can be safely done numerically.
Thus Eq.~(\ref{left4}) is a useful alternative representation for the
spectral density. In practice the integration over the semi-circles is 
done by expanding the integrand in $z$ for small $z$ and keeping only 
terms singular in $\epsilon$. The expansion requires only a finite number 
of terms and is a purely algebraic operation. Then the singularity in 
$\epsilon$ exactly cancels against those of the remaining integrals. This 
cancellation can also be done analytically leaving well defined and smooth 
integrands for further numerical treatment.

Next we turn to the threshold behaviour of the spectral density. Using the
results of the above analysis one finds 
\begin{eqnarray}\label{fin}
2m\rho(m^2)\sim\int^\infty K_1(mr)\prod_iZ_1(m_ir)dr
\end{eqnarray}
where $Z_1(m_ir)$ is either $I_1(m_ir)$ or $K_1(m_ir)$. The convergence at 
large $r>0$ is controlled by the factor $\exp{(-(m-M)r)}$ as in 
Eq.~(\ref{3final}), and the corresponding expansions in the variable $m-M$ 
in the region $m\sim M$ can be easily constructed.

To summarize, we have described a novel technique to compute the 
spectral density of any water melon diagram which reduces the calculation 
to a one-dimensional integral with well known functions in the integrand. 
Any tensor structure and form factor structure can be easily added without 
modification of the basic formulae. Different regimes of behavior with 
respect to mass/momentum expansions can easily be written down in terms of 
only two dimensionful parameters (one mass and one momentum or two masses) 
while other parameters are taken as small. The threshold behaviour of the 
spectral density can be easily investigated based on such a representation.
Explicit analytical formulae (with even that last integration being 
performed) are found in odd space-time dimensions. This allows one to 
compute any $n$-particle phase space integral for any kind of particles. 
The only reason that prevents the final integration to be explicitly done 
in even number of dimensions is that products of Bessel functions are 
encountered which cannot be integrated in closed form. This forces one to 
use numerical integrations in the even-dimensional case except for very 
special cases. Even then the analytic structure of the solution is 
completely determined which allows for an efficient and reliable numerical 
treatment of the problem.

All in all we conclude that the problem of computing the class of 
multi-loop massive water melon diagrams can be considered to be completely 
solved.

\vspace{1truecm}\noindent
{\large\bf Acknowledgements}\\[2mm]
We would like to thank A. Davydychev for useful discussions.
The work is supported in part by the Volkswagen Foundation under contract
No.\ I/73611. A.A.~Pivovarov is supported in part by the Russian Fund for 
Basic Research under contracts No.~96-01-01860 and 97-02-17065.


\begin{thebibliography}{99}
\bibitem{Gasser}J.~Gasser and M.E.~Sainio, Eur.~Phys.~J.\ {\bf C6} (1999) 297
\bibitem{PS}P.~Post and K.~Schilcher, Phys.~Rev.~Lett.\ {\bf 79} (1997) 4088
\bibitem{PT}P.~Post and J.B.~Tausk, Mod.~Phys.~Lett.\ {\bf A11} (1996) 2115
\bibitem{davyd}F.A.~Berends, A.I.~Davydychev and N.I.~Ussyukina,
  Phys.~Lett.\ {\bf 426 B} (1998) 95
\bibitem{ChKuhnKw}K.G.~Chetyrkin, J.H.~K\"uhn and A.~Kwiatkowski,
  Phys.~Rep.\ {\bf 277} (1996) 189  
\bibitem{Xloops}L.~Br\"ucher, J.~Franzkowski and D.~Kreimer,\\
  Report No.~MZ-TH-97-35, hep-ph/9710484 
\bibitem{GKP}S.~Groote, J.G.~K\"orner and A.A.~Pivovarov,
  Nucl.~Phys.\ {\bf B542} (1999) 515
\bibitem{GR}I.S.~Gradstein, I.M.~Ryzhik, ``Tables of integrals, series,
  and products'',\\ Academic Press, 1994
\bibitem{Watson}G.N.~Watson, ``Theory of Bessel functions'', Cambridge 1944.
\bibitem{Sch}A.A.~Pivovarov, N.N.~Tavkhelidze and V.F.~Tokarev,\\
  Phys.~Lett.\ {\bf 132 B} (1983) 402
\bibitem{fact}K.G.~Chetyrkin and A.A.~Pivovarov,
  Nuovo Cim.\ {\bf 100 A} (1988) 899
\bibitem{Sch1}A.A.~Pivovarov and V.F.~Tokarev, 
  Yad.~Fiz.\ {\bf 41} (1985) 524
\bibitem{Ktr0}C.S.~Meijer, Proc.~Amsterdam Akad.~Wet. (1940) 599; 702
\bibitem{Ktr}A.~Erdelyi (editor), ``Tables of integral transformations'',\\
  Volume~2, Bateman manuscript project, 1954.
\bibitem{NarPiv}S.~Narison and A.A.~Pivovarov, 
  Phys.~Lett.\ {\bf B327} (1994) 341
\bibitem{Mirkes}E.~Mirkes, Report No.~TTP-97-39, hep-ph/9711224
\bibitem{qcd30}D.J.~Gross, R.D.~Pisarski and L.G.~Yaffe,
  Rev.~Mod.~Phys.\ {\bf 53} (1981) 43
\bibitem{qcd3}A.M.~Polyakov, Phys.~Lett.\ {\bf 72 B} (1978) 477
\bibitem{Hatsuda}T.~Hatsuda, Nucl.~Phys.\ {\bf A544} (1992) 27
\bibitem{Raj}A.K.~Rajantie, Nucl.~Phys.\ {\bf B480} (1996) 729 
\bibitem{BS}N.N.~Bogoliubov and D.V.~Shirkov, ``Quantum fields'', 
  Benjamin, 1983.
\end{thebibliography}
\end{document}